\documentclass{sbrt2017eng}
\usepackage{mathrsfs,amsmath}
\usepackage{amsfonts}
\usepackage{graphicx}
\usepackage{hyperref}
\usepackage{subcaption}
\newcommand{\doi}{doi:}
\renewcommand*{\doi}[1]{\href{http://dx.doi.org/#1}{doi: #1}}
\begin{document}

\title{von Mises Tapering: A Circular Data Windowing}

\author{H. M. de Oliveira and F. Chaves
\thanks{H. M. de Oliveira, qPGOM, Statistics Department, Federal University of Pernambuco (UFPE), Recife-PE, Brazil, E-mail: hmo@de.ufpe.br. F. Chaves, Mechanics Department, Federal University of Pernambuco (UFPE), Recife-PE, Brazil, E-mail: chicouchaves@gmail.com} }

\maketitle

\begin{abstract}
Continuous standard windowing is revisited and a new taper shape is introduced, which is based on the normal circular distribution by von Mises. Continuous-time windows are considered and their spectra obtained. A brief comparison with  classical window families is performed in terms of their spectral properties. These windows can be used as an alternative in spectral analysis.
\end{abstract}

\section{Introduction}
Due to the cyclic or quasi-periodic nature of various types of signals, the signal processing techniques developed for real variables in the real line may not be appropriate. For circular data \cite{Berens}, \cite{Damien}, it makes no sense to use the sample mean, usually adopted for the data line as a measure of centrality. Circular measurements occur in many areas \cite{Jammalamadaka}, such as biology (or Chronobiology) \cite{Karp-Boss}, economy \cite{Dalkir}, \cite{Clark}, geography \cite{Clark_Burt}, medical (Circadian therapy \cite{Kirst}, epidemiology \cite{Gao}...), geology \cite{Watson}, \cite{Prevot}, \cite{Masuda}, meteorology \cite{Buishand}, \cite{Coles}, \cite{Ringelband}, acoustic scatter \cite{Jenison} and particularly in signals with some cyclic structure (GPS navigation \cite{Luo}, characterization of oriented textures \cite{Costa}, \cite{Peron}, discrete-time signal processing and over finite fields). Even in political analysis \cite{Gill}. Probability distributions can be successfully used as a support tool for several purposes: for example, the beta distribution was used in wavelet construction \cite{de_Oliveira}. In this paper, the von Mises distribution (VM) is used in the design of tapers. Tools such as rose diagram \cite{Mardia}, \cite{Izbicki} allow rich graphical interpretation. For random signals, the focus is on circular distributions of probability. The uniform distribution of an angle $\phi$, circular in the range $[0,2\pi]$, is given by
\begin{equation}
f_{\Phi;1}(\phi):=\frac{1}{2 \pi}\mathbb{I}_{[0,2\pi]}(\phi),
\end{equation}
where $\mathbb{I}_A(.)$ is the indicator function of the interval $A \subset \mathbb{R}$. It is denoted by $\phi \sim U(0,2\pi)$.\\
Another very relevant circular distribution is the normal circular distribution, introduced in 1918 by von Mises \cite{Von_Mises}, defined in the interval $[0,2\pi]$ and denoted by $\phi \sim VM(\phi_0,\beta)$,
\begin{equation}
f_{\Phi;2}(\phi):=\frac{1}{2 \pi I_0(\beta)}e^{\beta \cos(\phi - \phi_0)},
\end{equation}
where $\beta \geq 0$ and $I_0(.)$ is the zero-order modified Bessel function of the first kind \cite{Abramowitz}, \cite{Hill} (not to be confused with the indicator function), i.e. 
\begin{equation}
I_0(z):=\frac{1}{\pi} \int_{0}^{\pi}e^{z \cos\theta}d\theta=\sum_{n=0}^{+\infty}\frac{(z/2)^{2n}}{{n!}^2}.
\end{equation}
\begin{figure}[!ht]
\centering
\includegraphics[width=8cm]{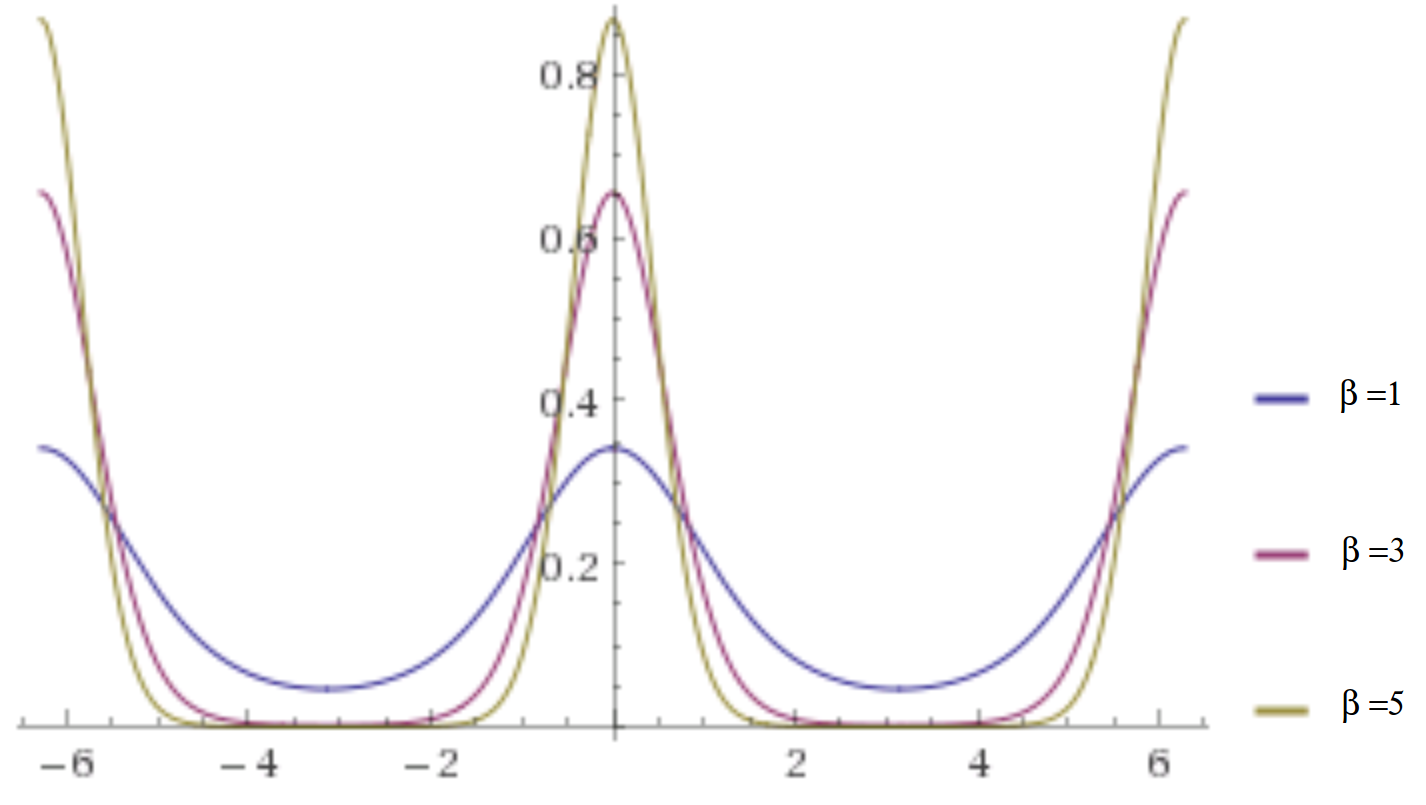}
\caption{{Periodic extension of the von Mises distribution with zero-mean for several parameter values:  $\beta=$ 1,3,5. Note that the support of the density is confined to $[-\pi,\pi]$.}}
\label{fig:vonMises}
\end{figure}
This probability density dominates in current analysis of circular data because it is flexible with regard to the effect of parameters and easy to interpret. In a standard notation,
\begin{equation}
f(x| \mu, \kappa):= \frac{e^{\kappa \cos(x-\mu)}}{2\pi I_0(\kappa)} \mathbb{I}_{[-\pi,\pi]}.
\end{equation}
Standardized distribution support is $[-\pi,\pi]$ and the mean, mode and median values are equal to $\mu$. The parameter $\kappa$ plays a role connected to the variance, namely $\sigma ^2 \approx 1/\kappa$.
\begin{equation*}
\mathbb{E}(X)=\mu ~\textnormal{and}~\mathbb{V}ar(X)=1-\frac{I_1(\kappa)}{I_0(\kappa)}.
\end{equation*}
Two limiting behaviors can be observed:
\begin{itemize}
\item
\begin{equation}
\lim_{\kappa \rightarrow 0} f(x|\mu,\kappa)=\frac{1}{2\pi}\text{rect} \left ( \frac{x}{2\pi} \right ),
\end{equation}
where $\text{rect}(x):= 
\left\{\begin{matrix}
1, & \textnormal{if }|x| \leq 1/2\\ 
0, & \textnormal{otherwise.}
\end{matrix}\right.$ is the normalized gate function, and therefore 
\begin{equation}
\lim_{\kappa \rightarrow 0} VM(\mu,\kappa) \sim U(-\pi,\pi).
\end{equation}
\item
\begin{equation}
\lim_{\kappa \rightarrow +\infty } f(x|\mu,\kappa)=\frac{1}{\sqrt{2 \pi \sigma ^2}} e^{-\frac{(x-\mu)^2}{2 \sigma ^2}},
\end{equation}
where $\sigma ^2:=1/\kappa$, and therefore
\begin{equation}
\lim_{\kappa \rightarrow +\infty} VM(\mu,\kappa) \sim N(\mu,1/\kappa).
\end{equation}
\end{itemize}
Hence the reason why this distribution is also known as the \textit{circular normal distribution}. The von Mises distribution (VM) is considered to be a circular distribution having two parameters and is the natural analogue on the circle of the Normal distribution on the real line. Maximum entropy distributions are outstanding probability distributions, because maximizing entropy minimizes the amount of prior information built into the distribution. Furthermore, many physical systems tend to move towards maximal entropy configurations over time. This encompasses distributions such as uniform, normal, exponential, beta... 
\begin{figure}[!hbt]
\centering
\includegraphics[width=8 cm, height=4 cm]{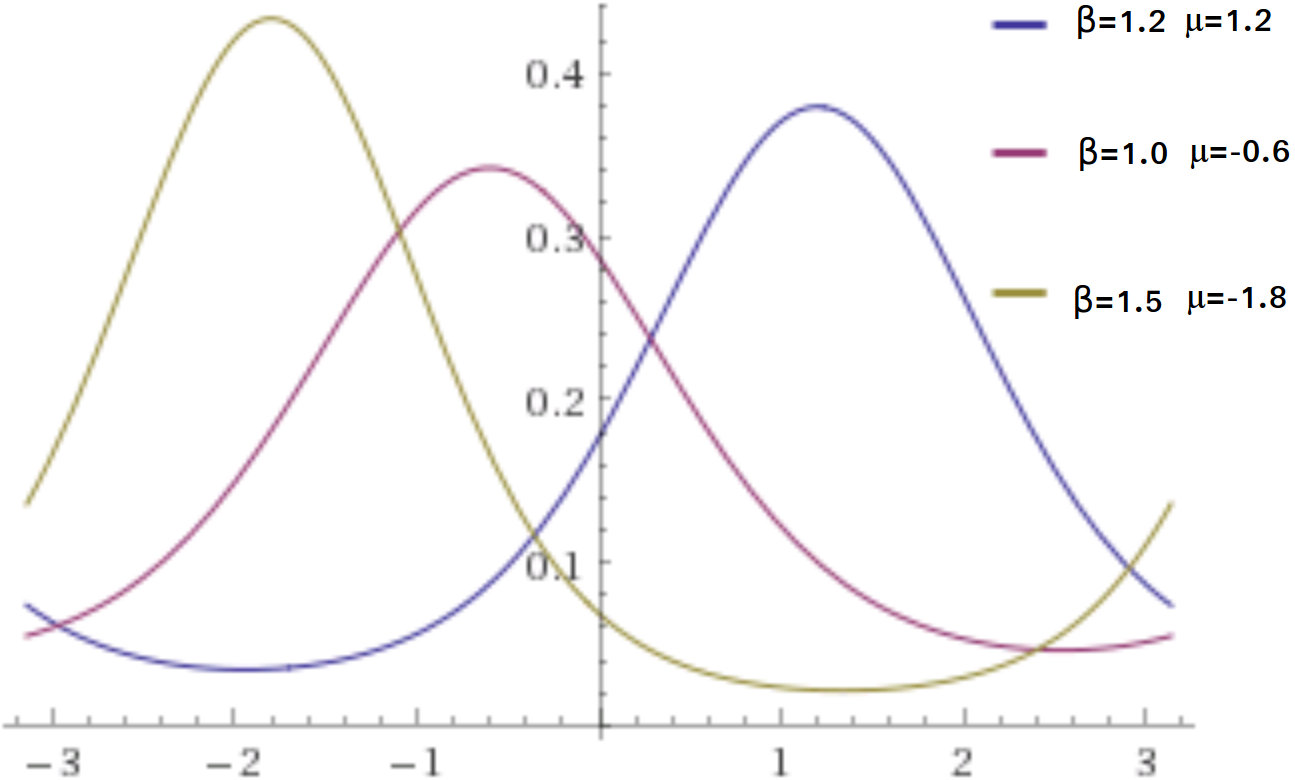}
\caption{{Circular behavior of the von Mises distribution plotted for different mean values (1.2, -0.6 and -1.8). The cyclical feature of the distribution is explained, in this case, within $[-\pi,\pi]$.}}
\label{fig:ciclicvonMises}
\end{figure}
\\
The von Mises distribution is the maximum entropy distribution for circular data when the first circular moment is specified \cite{Jammalamadaka}. The corresponding cumulative distribution function (CDF) is expressed by
\begin{equation}
F_X(x| \mu,\kappa)=\frac{1}{2 \pi}\sum_{n=-\infty }^{+\infty }\frac{I_{|n|}(\kappa)}{I_0(\kappa)}\left ( x-|n| \right ) \text{Sa}\left ( n(x-\mu) \right ),
\end{equation}
where $\text{Sa}(x):=\sin(x)/x$ is the sample function \cite{Lathi}.\\
Through a simple transformation of random variable, the distribution support can be modified to an interval defined between two integers:
\begin{equation}
f_{X_1}(x):=\frac{e^{\beta \cos \left ( \frac{2\pi}{N}x \right ) }}{NI_0(\beta)},~\textnormal{circular in } 0\leq x\leq N.
\end{equation}
Another closely related continuous distribution (with a minimal - but relevant difference) is
\begin{equation}
f_{X_2}(x):=\frac{e^{\beta \cos \left ( \frac{\pi}{N}x \right ) }}{NI_0(\beta)},~\textnormal{circular in } 0\leq x\leq N.
\end{equation}
This distribution has a circular pattern as best illustrated in Figure ~\ref{fig:ciclicvonMises}. Decaying pulses for constraining the signal support play a key role in a large number of domains, including: tapers \cite{Durrani} \cite{Parker}, linear networks (filtering \cite{Hayes}, inter-symbolic interference control \cite{Lathi}, modulation), wavelets \cite{deOliveira}, time series, Fourier transform spectroscopy \cite{Naylor} ... 
\section{Tapering: Standard Windows}
\begin{figure}[!ht]
\centering
{\includegraphics[width=4.2cm]{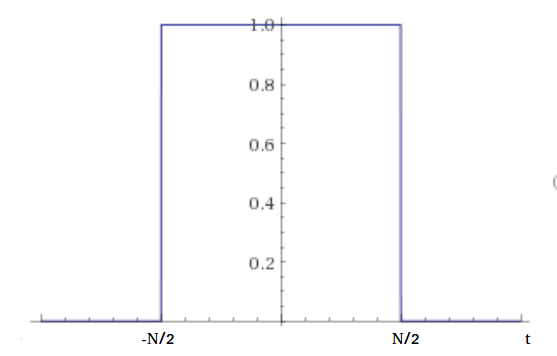}}
{\includegraphics[width=4.2cm]{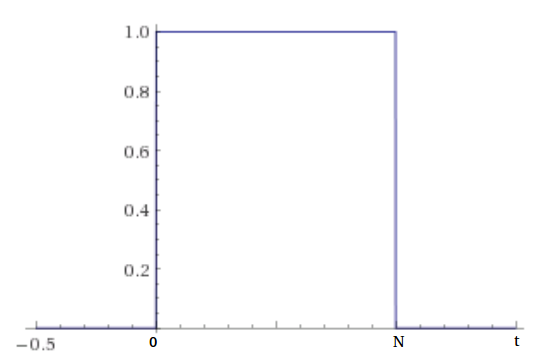}}
\caption{rectangular windows with length $N$. a) continuous, b) continuous causal. }
\label{fig:rect_windows}
\end{figure}

In this section we review some of the continuous windows (also known as a apodization function) used in signal processing (spectrum analysis \cite{Nuttall}, \cite{Durrani}), antenna array design \cite{Van_Veen},  characterization of oriented textures \cite{Peron}, image warping (\cite{Hayes}). Although discrete windows are more common, some studies address continuous windows \cite{Theubl}, \cite{Geckinli}, besides their application in short-time Fourier transforms. Among the most used windows, it is worth mentioning: rectangular, Bartlett, cosine-tip, Hamming, Hanning, Blackman, Lanczos, Kaiser, modified Kaiser, de la Vallé-Pousin, Poisson, Saram\"aki \cite{Saramaki}, \textit{Dolph-Chebyshev...} (non-exhaustive list \cite{Poularikas}). Tutorials on the subject are available \cite{Gautam}, \cite{Poularikas}, \cite{Harris}, \cite{Gopika}. Figure ~\ref{fig:rect_windows} describes the approaches to the standard rectangular window in two cases: 1) continuous non-causal, 2) continuous causal.\\
It is worth revisiting the spectra of each of these windows.
\begin{subequations}
\begin{align}
w_{REC;1}(t)&={\text{rect}}\left ( \frac{t}{N} \right ),\label{a}\\
w_{REC;2}(t)&={\text{rect}}\left ( \frac{t-N/2}{N} \right ).\label{b}
\end{align}
\end{subequations}
In the continuous case, $w_1(t)$ has spectrum given by:
\begin{equation}
W(w):=\mathscr{F}[w(t)]=\int_{-\infty }^{+\infty }w(t)e^{-jwt}dt.
\label{eq:ContinuoustimeFOURIER}
\end{equation}
Indeed $W_{REC;1}(w)=N \text{Sa}\left ( \frac{wN}{2}  \right )$. Now the spectrum of $w_{REC;2}(t)=\text{rect}\left ( \frac{t-N/2}{N}  \right )$ can be evaluated using the time-shift theorem \cite{Lathi}, $w(t-t_0)\leftrightarrow W(w) e^{-jwt_0}$, resulting in $W_{REC;2}(w)=N \text{Sa}\left ( \frac{Nw}{2}  \right )e^{-jNw/2}$.\\
\\
Several of the windows of interest can be encompassed taking into account the following definition
\begin{equation}
w_{\alpha;1}(t):= \left \{ \alpha+(1-\alpha) \cos\left ( \frac{2\pi}{N}t\right ) \right \} \text{rect}\left ( \frac{t}{N} \right ),
\label{eq:general_alpha}
\end{equation}
The Hanning (Raised Cosine) window corresponds to $\alpha=0.5$, whereas the standard Hamming window corresponds to  $\alpha=0.54$ \cite{Podder}. In the case of a cosine-tip continuous window ($\alpha=0$), the corresponding window and spectrum are \cite{Geckinli}
\begin{equation}
w_{\alpha=0;1}:=\cos\left ( \frac{2\pi}{N}t\right ) \text{rect}\left ( \frac{t}{N} \right ),
\end{equation}
and from the convolution theorem in the frequency \cite{Lathi})
\begin{equation}
W_{\alpha=0;1}(w)=\frac{N}{2} \text{Sa}\left ( \frac{Nw}{2}-\pi  \right )+ \frac{N}{2} \text{Sa}\left ( \frac{Nw}{2} +\pi \right ).
\end{equation}
The Kaiser window in continuous variable is defined by (non-causal window centered on the origin, and its corresponding causal version)
\begin{subequations}
\begin{align}
w_{KAI;1}(t):=&\frac{I_0\left ( \beta\sqrt{1-\left ( \frac{t}{N/2} \right )^2} \right )}{I_0(\beta))} \text{rect}\left ( \frac{t}{N} \right ),\label{a}\\
w_{KAI;2}(t):=&
\frac{I_0\left ( \beta\sqrt{1-\left ( \frac{t-N/2}{N/2} \right )^2} \right )}{I_0(\beta))} \text{rect}\left ( \frac{t-N/2}{N} \right ).\label{b}
\end{align}
\end{subequations}
The corresponding spectrum is given by:
\begin{equation}
W_{KAI;1}(w)=\frac{N}{I_0(\beta)} \text{Sa}\left ( \sqrt{(\frac{Nw}{2})^2 - \beta ^2} \right ).
\end{equation}
\section{Introducing a New Window: the Circular Normal Window}
\begin{figure}[ht]
\centering
\begin{subfigure}{0.45\textwidth}
\includegraphics[width=8 cm, height=3.5cm]{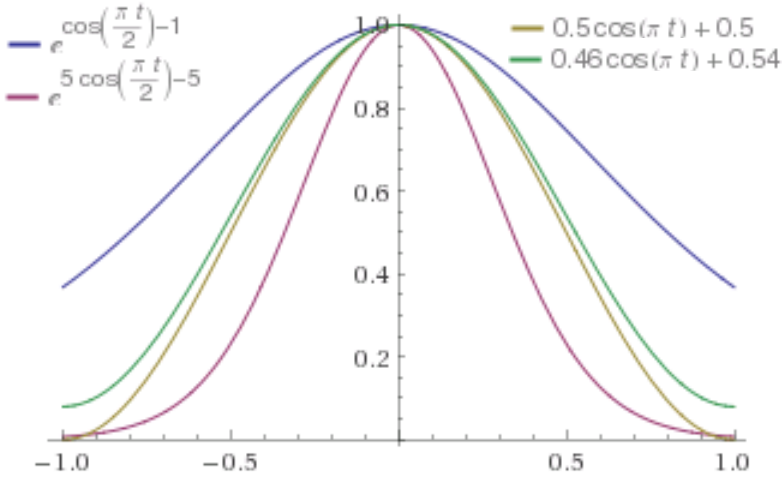}
\captionof{figure}{\small{Windows shape: von Hann, Hamming, circular $\beta=1,~5$.}}
\end{subfigure}%
\hfill
\begin{subfigure}{0.45\textwidth}
\centering
\includegraphics[width=8cm, height=3.8cm]{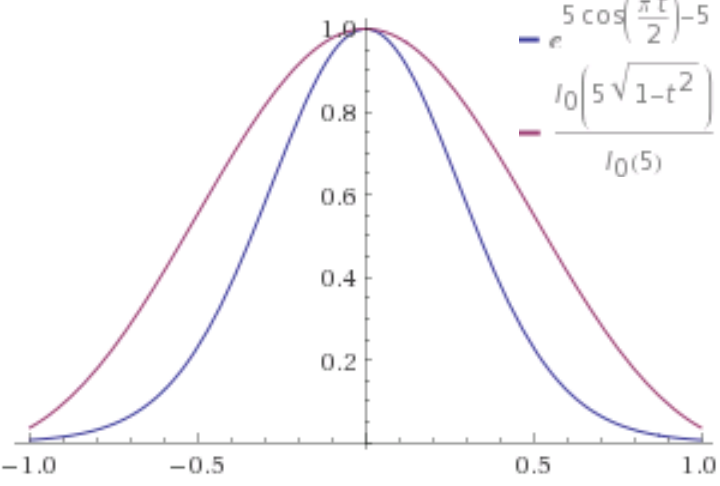}
\captionof{figure}{\small{Windows shape: Kaiser vs normal circular windows, $\beta=5$.}}
\end{subfigure}%
\caption{Shape comparison of different normalized windows for support $[-1,1]$.}
\label{fig:comparison_Han_Ham_CIR}
\end{figure}

The proposal is to use a window (support length $N$)  with shape related to the function (This type of windowing is superficially mentioned for spatial smoothing of spherical data \cite{Khalid})
\begin{equation}
w(t)=K \frac{e^{\beta \cos\left ( \frac{\pi}{N}t \right )}} {I_0(\beta)} \text{rect}\left ( \frac{t}{N} \right ).
\end{equation}
The value of the constant $K$ can be set so that, as in the other classic windows, $w(0)=1$. Thus, for continuous case (noncausal and causal, respectively), one has
\begin{subequations}
\begin{align}
w_{CIR;1}(t)&=\frac{e^{\beta \cos\left ( \frac{\pi}{N}t \right )}} {e^\beta} \text{rect}\left ( \frac{t}{N} \right ),\label{a}\\
w_{CIR;2}(t)&=\frac{e^{\beta \cos\left ( \frac{\pi}{N}t \right )}} {e^\beta} \text{rect}\left ( \frac{t-N/2}{N} \right ).\label{b}
\end{align}
\end{subequations}
A straightforward comparison among different tapers in displayed in Figure ~\ref{fig:comparison_Han_Ham_CIR}.

\section{Spectrum calculation of the Normal Circular Window: the Continuous Case}
\begin{figure}[!ht]
\centering
\includegraphics[width=6cm, height=4cm]{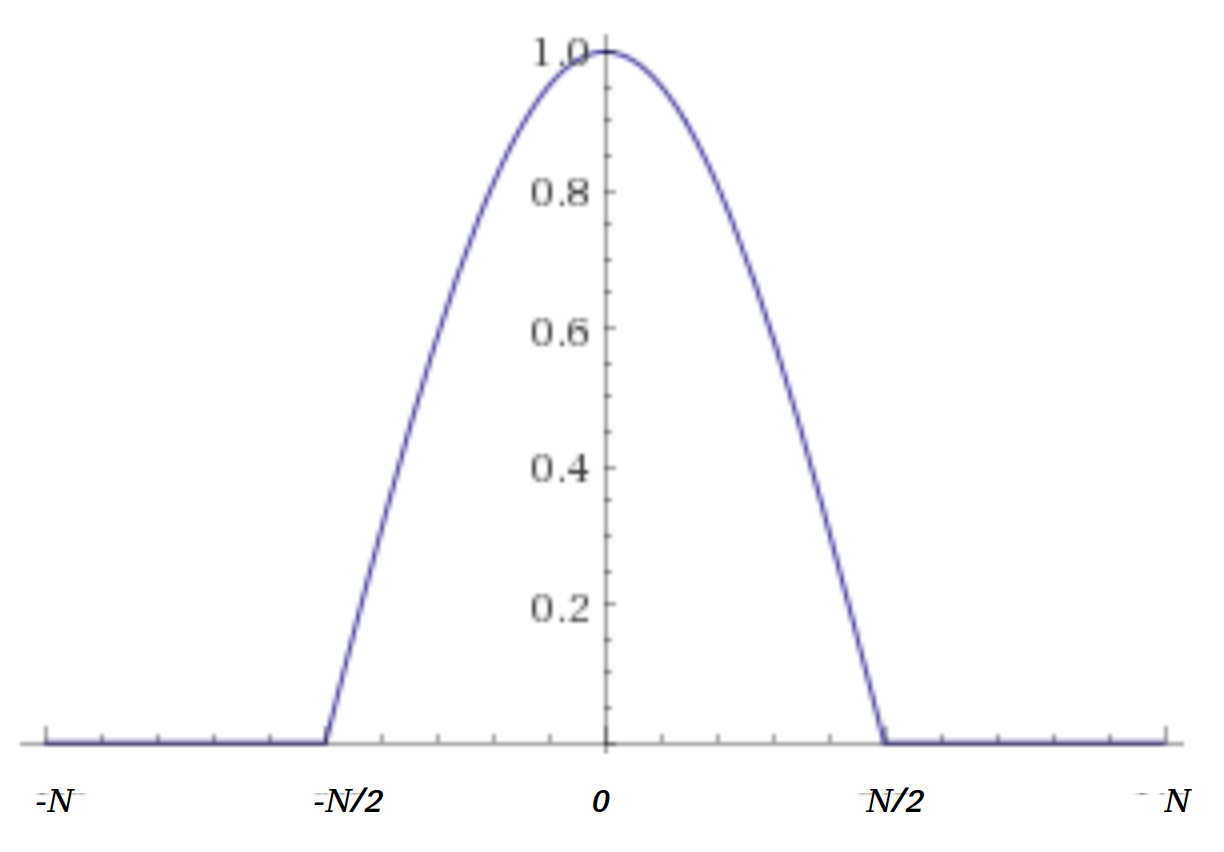}
\caption{Normalized cosine exponent of the exponential function in von Mises window: the (entire) cosine $\cos(\pi t/N)$ is periodic in $[-N,N]$, but the support is confined within $[-N/2,N/2]$ due to the rectangular pulse.}
\label{fig:cut_by_gate_window}
\end{figure}

In order to evaluate the spectrum of the continuous window introduced in the previous section, we use Eq.~\eqref{eq:ContinuoustimeFOURIER},
\begin{equation}
W_{CIR;1}(w)=\int_{-N/2}^{N/2}e^{\beta \left [ \cos \left ( \frac{\pi}{N}t \right )-1 \right ]}e^{-jwt}dt.
\end{equation}
The interest function involved in defining the window is $\cos \left ( \frac{\pi}{N}t \right )$, with period $2N$, sketched below in $[-N,N]$.  The rectangular term included in the window is responsible for cutting the window, confining it in the range $[-N/2,N/2]$ as viewed in Figure ~\ref{fig:cut_by_gate_window}. 
MacLaurin's series development of $e^{\beta.\left [ \cos \left ( \frac{\pi}{N}t \right ) \right ]}$ gives the following Fourier series:
\begin{equation}
e^{\beta \left [ \cos \left ( \frac{\pi}{N}t \right ) \right ]}=\sum_{n=-\infty }^{+\infty }I_{|n|}(\beta)
\cos\left (  \frac{n \pi}{N}t \right ).
\end{equation}
Thus, one obtains:
\begin{equation}
W_{CIR;1}(w)=e^{-\beta}\sum_{n=-\infty }^{+\infty }I_{|n|}(\beta) 
\mathscr{F}~\left ( \cos\left (  \frac{n \pi}{N}t \right ) \text{rect} \left ( \frac{t}{N} \right ) \right ).
\end{equation}
From the convolution theorem in the frequency \cite{Lathi}), the spectrum sought (a bit detailed here) is
\begin{flalign*}
W_{CIR;1}(w)= 
\end{flalign*}
\begin{equation}
=\frac{1}{2\pi}e^{-\beta}\sum_{n=-\infty }^{+\infty }I_{|n|}(\beta) 
\mathscr{F}~\left ( \cos\left (  \frac{n \pi}{N}t \right )\right )* \mathscr{F} ~\left ( \text{rect} \left ( \frac{t}{N} \right ) \right ).
\end{equation}
By evaluating the internal terms in the summation, we obtain
\begin{equation}
\frac{N}{2\pi}\pi \left \{ \delta \left ( w-\frac{n\pi}{N} \right )+\delta \left ( w+\frac{n\pi}{N} \right ) \right \}*\text{Sa}\left ( \frac{wN}{2} \right ),
\end{equation}
where $\delta$(.) is the Dirac impulse \cite{Lathi}, so that
\begin{equation*}
W_{CIR;1}(w)=\frac{N}{2} e^{-\beta}.
\end{equation*}
\begin{equation}
\sum_{n=-\infty }^{\infty } I_{|n|}(\beta)
\left \{ \text{Sa} \frac{N}{2} \left ( w-\frac{n\pi}{N} \right ) + \text{Sa} \frac{N}{2} \left ( w+\frac{n\pi}{N} \right )\right \},
\end{equation}
or
\begin{equation}
W_{CIR;1}(w)=Ne^{-\beta}\sum_{n=-\infty }^{\infty } I_{|n|}(\beta) \left \{ \text{Sa} \left ( \frac{Nw}{2}-\frac{n\pi}{2} \right ) \right \}.
\label{eq:series}
\end{equation}
This expression is like a series of reconstitution (with coefficients $c_n$)  of the type
\begin{equation*}
\sum_{n=-\infty }^{+\infty }c_n~ \text{Sa}\left ( \frac{Nw}{2}-\frac{n\pi}{2} \right ).
\end{equation*}
Let us apply the Shannon-Nyquist-Koteln'kov sampling theorem in the frequency domain, for time-limited signals (\cite{Luke}, \cite{Jerri}, \url{http://ict.open.ac.uk/classics}.\\
Since 
\begin{equation}
F(w)=\frac{w_st_m}{\pi}\sum_{n=-\infty }^{+\infty }F(nw_s) \text{Sa} \left ( wt_m-nt_mw_s \right ).
\end{equation}
The rate $w_s$ must comply with the restriction $w_s  \leq \pi/t_m$, and the choice made is $w_s= \pi/2t_m$, so that the previous equation is
\begin{equation}
F(w)=\frac{1}{2}\sum_{n=-\infty }^{+\infty }F(\frac{n\pi}{2t_m})\text{Sa} \left (w t_m -\frac{n\pi}{2} \right ).
\end{equation}
Now let us choose the duration $t_m$ to be $t_m:=N/2$ (Figure ~\ref{fig:cut_by_gate_window}), which leads to
\begin{equation}
F(w)=\frac{1}{2}\sum_{n=-\infty }^{+\infty }F(\frac{n\pi}{N})\text{Sa} \left ( \frac{Nw}{2}-\frac{n\pi}{2} \right ).
\label{eq:reconstruction_n_pi_over_2}
\end{equation}  
This is a variation of the cardinal Whittaker-Shannon series \cite{Marks}. Observing the series described in Eq.~\eqref{eq:series}, it is seen that the signal corresponds to a continuous signal defined by samples such that $F\left ( \frac{n\pi}{N} \right )=2I_{|n|}(\beta)$ and
\begin{equation}
F(w)=2I_{\left | \frac{Nw}{\pi} \right |}(\beta),
\end{equation}
and the spectrum is given by
\begin{equation}
W_{CIR;1}(w)=\frac{2NI_{\left | \frac{Nw}{\pi} \right |}(\beta)}{e^ {\beta}}.
\end{equation}
In the case of the causal window, $w_{CIR;1}(t)$, the application of the time-shift theorem provides the spectrum
\begin{equation}
W_{CIR;2}(w)=\frac{2NI_{ \frac{N}{\pi} \left |w \right |}(\beta)}{e^ {\beta}}e^{-jw\frac{N}{2}}.
\end{equation}
It is worth remembering that the $\nu$ argument of the $I_{\nu}(z)$  function is a real number in this case \cite{Abramowitz}.

\section{Conclusions}
The closeness to the normal distribution and the fact that they are associated with a shape linked to the maximum entropy for circular data suggests interesting properties to be explored in later investigations. Windowing circular data with von Mises circular window can possibly improve spectral evaluation in these cases. Discrete data windows of this kind is currently under investigation.

\section*{Acknowledgements}
This work has been supported by Statistics Department UFPE, Brazil. HMdO thanks an anonymous referee for providing the reference \cite{Khalid}.


\begin{thebibliography}{99}

\bibitem{Abramowitz} Abramowitz, M., and I.A. Stegun.  \emph{Handbook of Mathematical Functions with Formulas, Graphs, and Mathematical Tables}. Vol.9. Dover, New York, 1972. \url{http://people.math.sfu.ca/~cbm/aands/abramowitz_and_stegun.pdf}.

\bibitem{Berens} Berens, P. CircStat: a MATLAB toolbox for circular statistics. \emph{J Stat Softw} 31.10: 1-21, 2009. \doi{10.18637/jss.v031.i10} 

\bibitem{Buishand} Buishand, T.A. Some methods for testing the homogeneity of rainfall records. \emph{Journal of hydrology} 58.1-2: 11-27, 1982. \doi{10.1016/0022-1694(82)90066-X}

\bibitem{Clark} Clark, W.A.V., Y. Huang, and S. Withers. Does commuting distance matter?: Commuting tolerance and residential change. \emph{Regional Science and Urban Economics} 33.2: 199-221, 2003 \doi{10.1016/S0166-0462(02)00012-1}.

\bibitem{Clark_Burt} Clark, W.A.V., and J.E. Burt. The impact of workplace on residential relocation. \emph{Annals of the Association of American Geographers} 70.1:59-66, 1980 \doi{10.1111/j.1467-8306.1980.tb01297.x}

\bibitem{Coles} Coles, S.G., and D. Walshaw. Directional modelling of extreme wind speeds. \emph{Applied Statistics}, 139-157, 1994
\doi{10.2307/2986118}.

\bibitem{Costa} Da Costa, J.-P, et al. Unsupervised segmentation based on Von Mises circular distributions for orientation estimation in textured images. \emph{Journal of Electronic Imaging} 21.2: 021102-1, 2012. \doi{10.1117/1.JEI.21.2.021102}. 

\bibitem{Dalkir} Dalkir, M. Revisiting stock market index correlations. \emph{Finance Research Letters} 6.1: 23-33, 2009 \doi{10.1016/j.frl.2008.11.004}.

\bibitem{Damien} Damien, P., and S. Walker. A full Bayesian analysis of circular data using the von Mises distribution. \emph{Canadian Journal of Statistics} 27.2: 291-298, 1999. \doi{10.2307/3315639}. 

\bibitem{deOliveira} de Oliveira, H.M., L.R. Soares, and T.H. Falk. A Family of Wavelets and a New Orthogonal Multiresolution Analysis Based on the Nyquist Criterion,  \emph{Journal of Communication and Information Systems} 18.1: 69-76, 2003. \doi{10.14209/jcis.2003.8}.

\bibitem{de_Oliveira} de Oliveira, H.M. and G.A.A. Araujo. Compactly Supported One-cyclic Wavelets Derived from Beta Distributions, Journal of Communication and Information Systems, vol.20, n.3, pp.27-33, 2005. \doi{10.14209/jcis.2005.17}

\bibitem{Durrani} Durrani, T.S., and J.M. Nightingale. Data windows for digital spectral analysis. \emph{Proceedings of the Institution of Electrical Engineers}. Vol. 119. No. 3. IET Digital Library, 1972. \doi{10.1049/piee.1972.0080}

\bibitem{Gao} Gao, F., et al. On the application of the von Mises distribution and angular regression methods to investigate the seasonality of disease onset. \emph{Statistics in medicine} 25.9: 1593-1618, 2006. \doi{10.1002/sim.2463}.

\bibitem{Gautam} Gautam, J.K., A. Kumar, and R. Saxena. Windows: A tool in signal processing. \emph{IETE Technical Review} 12.3: 217-226, 1995. \doi{10.1080/02564602.1995.11416530}.

\bibitem{Geckinli} Geçkinli, N., and D. Yavuz. Some novel windows and a concise tutorial comparison of window families. \emph{IEEE Transactions on Acoustics, Speech, and Signal Processing} 26.6: 501-507, 1978. \doi{10.1109/TASSP.1978.1163153}

\bibitem{Gill} Gill, J. and D. Hangartner. Circular data in political science and how to handle it, \emph{Political Analysis} 18(3): 316-336, 2010. \doi{10.1093/pan/mpq009}

\bibitem{Gopika} Gopika, P. FIR window method: A comparative analysis. \emph{IOSR Journal of Electronics and Communication Engineering}, Vol.10, N.4: 15-20, 2015.

\bibitem{Harris} Harris, F.J. On the use of Windows for harmonic analysis with the Discrete Fourier Transform. \emph{Proceedings of the IEEE} 66(1): 51-83. 1978. \doi{10.1109/PROC.1978.10837}.

\bibitem{Hayes} Hayes, M.H. \emph{Schaum’s Outline of Theory and Problems of Digital Signal Processing}, 1999, McGraw-Hill Companies [Chapter 9]

\bibitem{Hill} Hill, G.W. Algorithm 518: Incomplete Bessel function $I_0$. The Von Mises Distribution [S14]. \emph{ACM Transactions on Mathematical Software} (TOMS) 3.3: 279-284, 1977. \doi{10.1145/355744.355753}

\bibitem{Izbicki} Izbicki, R. \emph{Análise de Dados Circulares}, Trabalho de IC, Instituto de Matemática e Estatística da Universidade de São Paulo-IME-USP, 29p., 2008.

\bibitem{Jammalamadaka} Jammalamadaka, S. Rao, and A. Sengupta. \emph{Topics in circular statistics}. Vol. 5. World Scientific, 2001.

\bibitem{Jenison} Jenison, R. L., and K. Fissell. A comparison of the von Mises and Gaussian basis functions for approximating spherical acoustic scatter. \emph{IEEE transactions on neural networks} Vol 6.5: 1284-1287, 1995. \doi{10.1109/72.410375}.

\bibitem{Jerri} Jerri, A.J. The Shannon sampling theorem— Its various extensions and applications: A tutorial review.  \emph{Proceedings of the IEEE} 65.11 (1977): 1565-1596. \doi{10.1109/PROC.1977.10771}.

\bibitem{Kaiser} Kaiser, J., and R. Schafer. On the use of the $I_0$-sinh window for spectrum analysis.  \emph{IEEE Transactions on Acoustics, Speech, and Signal Processing} 28.1: 105-107, 1980. \doi{10.1109/TASSP.1980.1163349}.

\bibitem{Karp-Boss} Karp-Boss, L., E. Boss, and P.A. Jumars. Motion of dinoflagellates in a simple shear flow. \emph{Limnology and oceanography} 45.7: 1594-1602, 2000 \doi{10.4319/lo.2000.45.7.1594}.

\bibitem{Khalid} Khalid, Z., R. A. Kennedy, and S. Durrani. On the choice of window for spatial smoothing of spherical data. \emph{Acoustics, Speech and Signal Processing (ICASSP)}, 2014 IEEE International Conference on. IEEE, 2014.
\doi{10.1109/ICASSP.2014.6854079}

\bibitem{Kirst} Kirst, C., M. Timme, and D. Battaglia. Dynamic information routing in complex networks. \emph{Nature communications} 7, 2016. \doi{10.1038/ncomms11061}.

\bibitem{Lathi} Lathi, B.P. and Z. Ding. \emph{Modern Digital and Analog Communication Systems}, 4th Ed, Oxford University press, 2008.

\bibitem{Luke} Luke, H.D. The origins of the sampling theorem. \emph{IEEE Communications Magazine} 37.4: 106-108, 1999. \doi{10.1109/35.755459}

\bibitem{Luo} Luo, X., M. Mayer, and B. Heck. On the probability distribution of GNSS carrier phase observations.\emph{GPS solutions} 15.4: 369-379, 2011 \doi{10.1007/s10291-010-0196-2}

\bibitem{Mardia} Mardia, K.V., and P.J. Zemroch. Algorithm AS 86: The von Mises distribution function. \emph{Journal of the Royal Statistical Society}. Series C (Applied Statistics) 24.2: 268-272, 1975. \doi{10.2307/2346578}.

\bibitem{Marks} Marks II, R. \emph{Introduction to Shannon sampling and interpolation theory}. Springer Science \& Business Media, 2012.

\bibitem{Masuda} Masuda, T. et al. A statistical approach to determination of a mineral lineation. \emph{Journal of Structural Geology} 21.4: 467-472, 1999. \doi{10.1016/S0191-8141(99)00005-X}

\bibitem{Naylor} Naylor, D.A., and M.K. Tahic. Apodizing functions for Fourier transform spectroscopy. \emph{JOSA A} 24.11: 3644-3648, 2007. \doi{10.1364/JOSAA.24.003644}

\bibitem{Nuttall} Nuttall, A.H. Some windows with very good sidelobe behavior. \emph{IEEE Transactions on Acoustics, Speech, and Signal Processing} 29 (1): 84–91. 1981. \doi{10.1109/TASSP.1981.1163506}

\bibitem{Parker} Parker, K.J. Correspondence: apodization and windowing functions. \emph{IEEE transactions on ultrasonics, ferroelectrics, and frequency control} 60.6: 1263-1271, 2013. \doi{10.1109/TUFFC.2013.2691}.

\bibitem{Peron} Péron, M-C. et al. Modèles linéaires et circulaires pour la charactérisation de textures orientées. \emph{XXIIe colloque GRETSI} (traitement du signal et des images), Dijon (FRA), 8-11 septembre 2009.

\bibitem{Podder} Podder, P. et al. Comparative performance analysis of Hamming, Hanning and Blackman window. \emph{International Journal of Computer Applications} 96.18, 2014. \doi{10.5120/16891-6927}

\bibitem{Poularikas} Poularikas A.D. \emph{Windows: The Handbook of Formulas and Tables for Signal Processing}. Ed. A.D. Poularikas, Boca Raton: CRC Press LLC, 1999.

\bibitem{Prevot} Prévot, M., and P. Camps. Absence of preferred longitude sectors for poles from volcanic records of geomagnetic reversals. \emph{Nature} 366.6450:53-57,1993 \doi{10.1038/366053a0}.

\bibitem{Ringelband} Ringelband, T., P. Schäfer, and A. Moser. Probabilistic ampacity forecasting for overhead lines using weather forecast ensembles. \emph{Electrical Engineering} 95.2, 2013. \doi{10.1007/s00202-012-0244-8}
  
\bibitem{Saramaki} Saramaki, T. A class of window functions with nearly minimum sidelobe energy for designing FIR filters, 1989, \emph{IEEE International Symposium on Circuits and Systems}, 1989. \doi{10.1109/ISCAS.1989.100365}
 
\bibitem{Theubl} Theu{\ss}l, T., H. Hauser, and E. Gröller. Mastering windows: improving reconstruction. \emph{Proceedings of the 2000 IEEE symposium on Volume visualization}. ACM, 2000. \doi{10.1109/VV.2000.10002}

\bibitem{Van_Veen} Van Veen B.D. and K. Buckley. Beamforming: a versatile approach to spatial filtering, \emph{IEEE ASSP Magazine}, April: 4-24, 1988 \doi{10.1109/53.665}.

\bibitem{Von_Mises} Von Mises, R. \emph{Über die ‘Ganzzahligkeit’ der atomgewicht und verwandte Fragen}. Physikal. Z., Vol 19: 490–500, 1918.

\bibitem{Watson} Watson, G.S. The statistics of orientation data. \emph{The Journal of Geology} 74.5, Part 2: 786-797, 1966
 \doi{10.1086/627211}.

\end{thebibliography}
\end{document}